# Self-correcting longitudinal phase space in a multistage plasma accelerator

Carl A. Lindstrøm*

**Plasma accelerators driven by intense laser[1] or particle beams[2] provide gigavolt-per-meter accelerating fields[3,4], promising to drastically shrink particle accelerators for high-energy physics and photon science[5,6]. Applications such as linear colliders and free-electron lasers (FELs) require high energy and energy efficiency, but also high stability and beam quality. The latter includes low energy spread, which can be achieved by precise beam loading[7] of the plasma wakefield using longitudinally shaped bunches[8], resulting in efficient[9] and uniform[10] acceleration. However, the plasma wavelength, which sets the scale for the region of very large accelerating fields to be 100 μm or smaller, requires bunches to be synchronized and shaped with extreme temporal precision, typically on the femtosecond scale. Here, a self-correction mechanism is introduced, greatly reducing the susceptibility to jitter. Using multiple accelerating stages[11,12], each with a small bunch compression between them, almost any initial bunch, regardless of current profile or injection phase, will self-correct into the current profile that flattens the wakefield, damping the relative energy spread and any energy offsets. As a consequence, staging can be used not only to reach high energies, but also to produce the exquisite beam quality and stability required for a variety of applications.**

The specific requirements for energy spread and stability depend on the application. One of the most important limitations on the performance of a linear collider is the final-focusing system, which typically has an energy acceptance of less than 1%. Collider designs therefore specify a relative root-mean-square (rms) energy spread of around 0.2% (ILC[13]) to 0.35% (CLIC[14]), and a similar level of energy stability. In an FEL, the energy spread must be smaller than the Pierce parameter[15], typically ~0.1% rms or less. This applies mainly to the uncorrelated energy spread; the correlated energy spread can in principle be larger. While the lasing process itself does not require high energy stability, the optics delivering the beam to the lasing structure will typically have a percent-level energy acceptance. Similarly, injection into a storage ring also has an energy acceptance of ~1%.

Accelerating particle bunches in a plasma wakefield that varies on a timescale $1/\omega_p$, where $\omega_p$ is the plasma frequency, to within a relative precision $\sigma$ will require synchronization at the level $\sigma/\omega_p$. As an example, for a plasma density of $10^{16}$ cm$^{-3}$ and a required precision of 0.3%, this implies a tolerance of only 0.5 fs. For higher densities, the tolerance is even lower. Both the injection synchronization and the temporal resolution to which the current profile must be controlled will need to be of sub-fs order. Experimentally, this is very challenging, considering that state-of-the-art synchronization is approximately 10 fs[16].

Given these constraints, a correction mechanism that can loosen the required tolerances would be highly desirable. Within a single plasma-accelerator stage, however, the phase and current profile remains approximately constant, and causality prevents any feedback between the ultrarelativistic trailing bunch and the leading driver. Therefore, to correct relative energy errors within the bunch, the particles must be temporarily extracted such that the phase and current profile can be manipulated before acceleration in the next stage. This is the principle behind a proposal by Ferran Pousa et al.[17] to use a magnetic chicane with a fine-tuned $R_{56}$ (longitudinal lattice dispersion) between two identical stages, able to compensate for a linear chirp in one stage by exactly inverting the longitudinal phase space (LPS) before acceleration in the next. However, the nonlinear chirp introduced by beam loading reduces the effectiveness of this scheme. In this Letter, it is shown that by making use of *multiple* stages separated by magnetic chicanes, such beam loading facilitates a self-correction mechanism: a feedback between the shape of the current profile and the corresponding beam loading of the wakefield provides not only phase focusing[18,19], but also corrects the current profile in such a way that the acceleration becomes progressively more uniform.

To illustrate this self-correction mechanism, we consider an example with parameters relevant to a collider application. This requires modelling plasma accelerators with many stages. The large number of time steps needed results in high computational costs with a concomitant need for exascale computing[20], making a simplified model imperative. In fact, only the evolution in LPS is required for this problem. The longitudinal and transverse phase spaces are therefore treated as separate, assuming that the acceleration is independent of transverse position. This holds true in the ion column of a nonlinear plasma wake due to the Panofsky-Wenzel theorem[21,22]. It is also assumed that all the charge survives the coupling between stages and remains within the plasma-cavity structure. While these assumptions are idealized, they allow mechanisms in the LPS to be studied in isolation.

An analytical model developed by Lu et al.[23], and recently improved by Dalichaouch et al.[24], accurately describes beam loading in a nonlinear plasma wakefield[8], as verified by particle-in-cell (PIC) simulations. Extending this single-stage model to multiple stages separated by magnetic chicanes requires the introduction of a new, rudimentary PIC-like simulation with four distinct steps: (i) an initial LPS particle distribution is defined, as well as a 1D grid of longitudinal positions; (ii) inside the accelerator stage, the beam-loaded electric field in each grid position is calculated based on the current profile using the analytical model; (iii) the energy of each particle is then updated; (iv) outside the stage, the $R_{56}$ of the chicane is applied, shifting the longitudinal position of each particle based on its relative energy offset from the nominal chicane energy. Steps (ii)–(iv) are then repeated for every stage (see Methods).

In this example, starting at $E_0$ = 10 GeV, each stage accelerates the bunch by $\Delta E$ = 2 GeV up to a final energy of 500 GeV—a total of 245 stages. Each stage operates with a plasma density of $10^{16}$ cm$^{-3}$ and a blowout radius of $R_b = 2.5 k_p^{-1}$, where $k_p$ is the plasma wavenumber. The initial LPS distribution is bi-Gaussian with 2.3 nC of charge, $\sigma_{\delta,0} = 0.5\%$ rms initial energy spread, and a bunch length of 13 µm rms. The exact $R_{56}$-values of the magnetic chicanes are not critical for the mechanism—here it is modelled to start at $R_{56} = 0.34$ mm after the first stage, and then stage-to-stage scaled by the inverse square root of the energy[25]. Additionally, an uncorrelated energy spread of $\sigma_{\delta,\text{uncorr.}} = 0.5\%$ rms is imposed to take account of small transverse non-uniformities of the accelerating field (e.g., from ion motion).

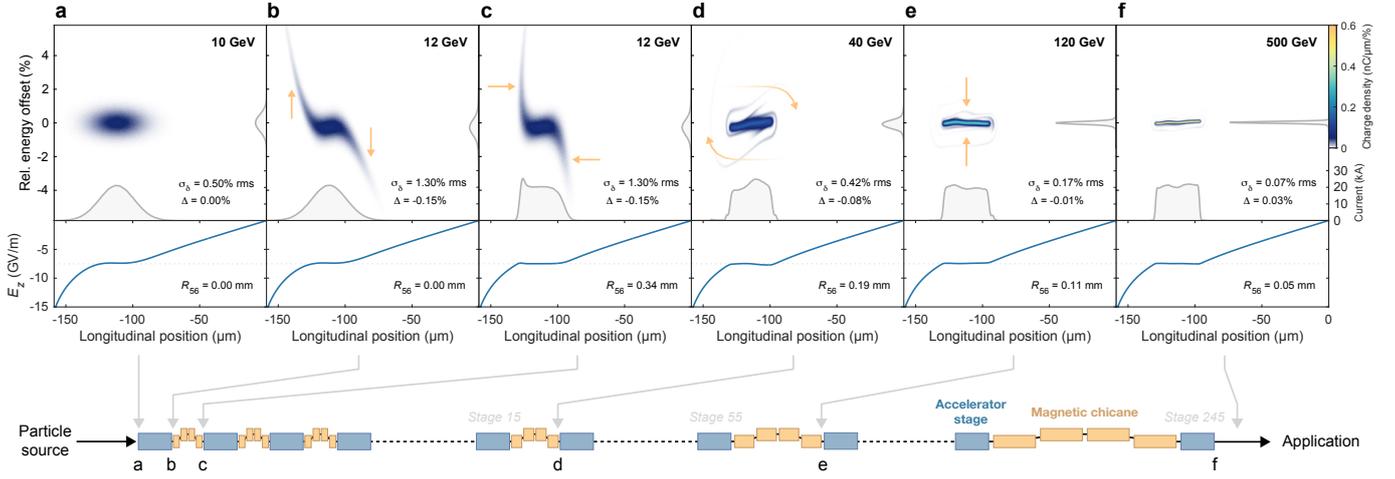

**Fig. 1 | Illustration of the self-correction mechanism. a,** An electron bunch, with an initial bivariate Gaussian LPS distribution, is injected into the first plasma-accelerator stage. **b,** Beam loading by the Gaussian current profile flattens the wakefield ($E_z$) near the bunch core, but results in energy errors for the bunch head and tail. **c,** A small compression ($R_{56}$) from a magnetic chicane between the stages shifts particles at higher energy forward and particles at lower energy backward, changing the current profile and therefore also the beam loading of the wakefield. **d,** This process of acceleration and compression repeats, gradually redistributing the charge to flatten the wakefield. **e,** While this redistribution occurs, the bunch gains energy stage-by-stage, leading to decreased relative energy offsets. **f,** The end result is an equilibrium (flattop) current profile that flattens the wakefield, and a significantly reduced relative energy spread.

Figure 1 shows the evolution of the accelerating bunch in LPS at representative locations throughout the accelerator. In the first stage, the nonuniform wakefield imparts a nonlinear chirp on the bunch. The positive $R_{56}$ of the chicane moves the position of lower-energy particles at the head backward and the higher-energy particles at the tail forward with respect to the core, which remains largely unperturbed. With every stage, this redistribution moves off-energy charge toward the center in an oval track around the bunch core, gradually flattening the wakefield. Since the absolute energy gain is identical for all particles at a given longitudinal position, regardless of initial energy, the relative energy offset of each particle will gradually decrease as the bunch accelerates. After a large number of stages, this *fold-and-squeeze* process reaches an equilibrium current profile and a very low relative energy spread. Further acceleration will continue to decrease the energy spread, approximately according to

$$\sigma_{\delta,i} = \frac{1}{E_i}\sqrt{E_0^2 \sigma_{\delta,0}^2 + \sigma_{E,\text{mismatch}}^2 + i \Delta E^2 \sigma_{\delta,\text{uncorr.}}^2}, \quad (1)$$

where $E_i$ is the energy after stage $i$, and $\sigma_{E,\text{mismatch}}$ is a growth in absolute energy spread due to a mismatch in LPS (from non-optimal beam loading).

The damping process, shown in Fig. 2, at first glance appears to violate Liouville's theorem. However, the area of the particle distribution in the *absolute* LPS (i.e., absolute energy versus longitudinal position) never decreases—in fact, this normalized longitudinal emittance even increases stage-to-stage. On the other hand, when viewed in the *relative* LPS (i.e., relative energy offset versus longitudinal position), most relevant to applications, the area of the particle distribution decreases with acceleration.

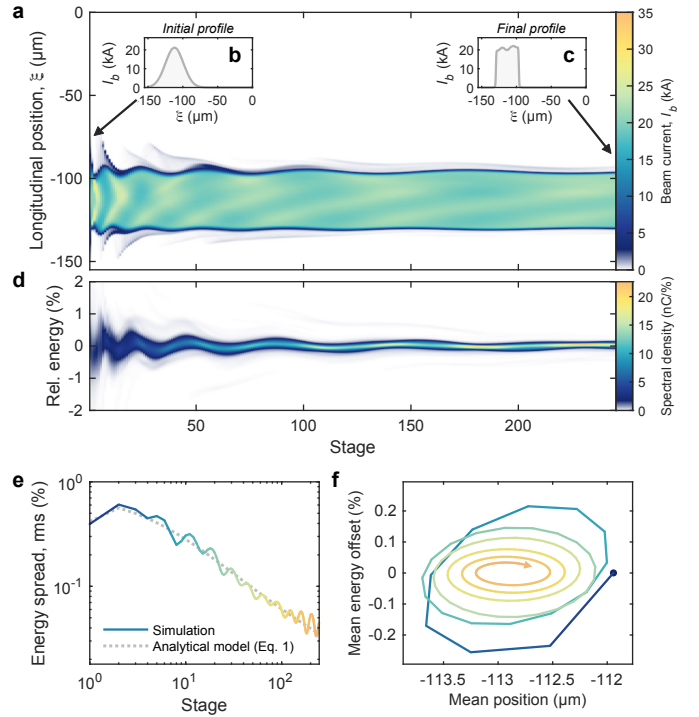

**Fig. 2 | Damping of energy spread and energy errors. a,** The evolution of the beam current, shown as a function of longitudinal position and stage number. **b,** The current profile starts as a Gaussian, gradually evolving to a flattop shape (**c**) across the 245 stages of the simulation. **d,** In this process, the relative energy spread and offset decrease. **e,** After an initial increase in energy spread due to a longitudinal-phase-space mismatch, the relative energy spread decreases with acceleration according to equation (1), where $\sigma_{E,\text{mismatch}} = 60$ MeV rms is used. **f,** At the same time, the bunch centroid exhibits an initial excursion in LPS (starting from the blue point), followed by synchrotron oscillations (at a rate proportional to $k_p R_{56}$) with a decaying amplitude. The line color corresponds to the stage number (logarithmically, as defined in **e**).

Insensitivity to the initial current profile, as seen in the example, is only one aspect of a general increase in tolerances, which also includes a considerably increased tolerance to timing jitter. Two forms of timing jitter can be considered: (i) the effect of timing jitter of the driver will average out over many stages, with or without the self-correction mechanism, and is therefore not a major concern unless this jitter is large enough for the accelerating bunch to miss the plasma cavity altogether; (ii) on the other hand, timing jitter of the initial injection phase (with respect to the average driver timing) can critically affect the final energy spectrum. Figure 3 shows how this sensitivity is greatly reduced when the self-correction mechanism is applied, using an example similar to that depicted in Figs. 1 and 2. Here, to allow also uncorrected bunches to be accelerated with per-mille-level energy spread, bunches are instead initialized with the optimal (flattop) current profile. In this case, the timing tolerance increases from ~1 fs to ~200 fs when the self-correction mechanism is applied, as defined by the span of injection timing able to deliver more than half of the initial charge within a final energy range of ±0.3%. This dramatic improvement places experimental realization of stable plasma acceleration well within the 10-fs limit of state-of-the-art synchronization.

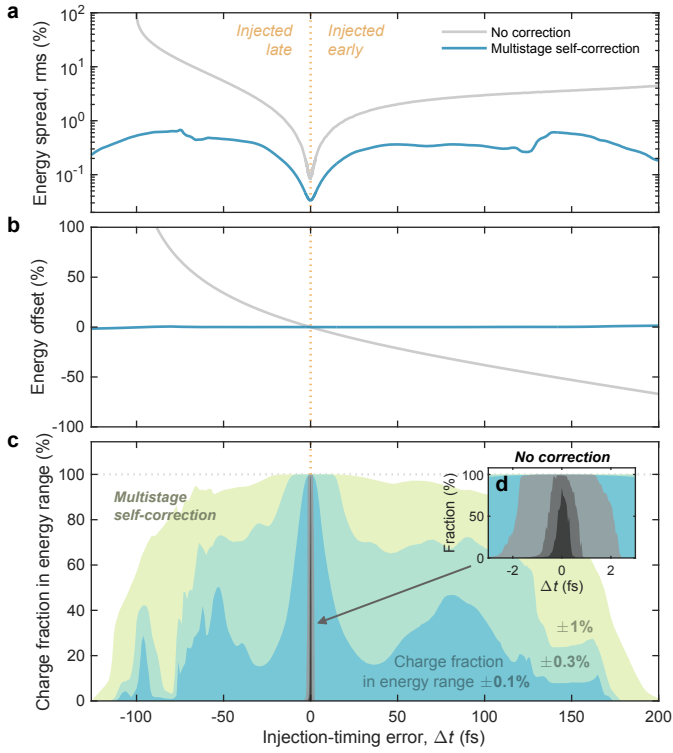
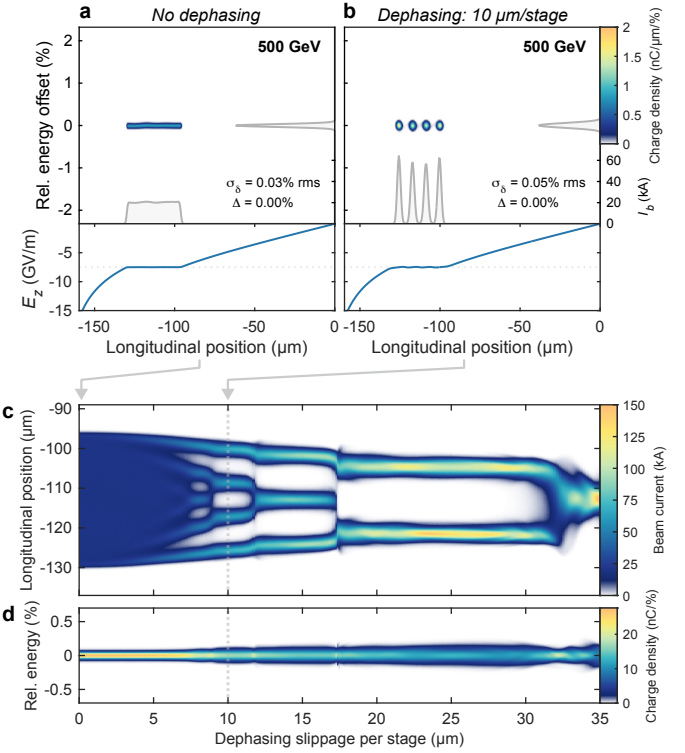

**Fig. 3 | Increased tolerance to errors in initial injection timing. a,** The final energy spread varies greatly with initial injection timing if not corrected (gray line), but remains consistently low when multistage correction is applied (blue line). **b,** Similarly, the final energy offset is highly sensitive to the initial timing error unless multistage self-correction is applied. **c,** Combining these, the timing tolerance for which a large fraction of charge is delivered within various spectral ranges (±0.1%, ±0.3% and ±1%) is shown to be significantly larger with self-correction (colored areas) compared to no correction (gray-shaded areas; zoomed in inset **d**).

**Fig. 4 | Self-correction in the presence of dephasing. a,** With no dephasing, the final and self-corrected current profile is a flattop. **b,** This changes when dephasing is introduced. By example, if a total slippage of 10 μm due to dephasing occurs in each of the 245 stages of the simulation, the averaged wakefield is best flattened by a train of four microbunches. **c,** Simulations with different amounts of dephasing reveal a bifurcation into microbunch trains, with sharp transitions between integer numbers of bunches. **d,** Since the wakefield cannot be fully flattened by the bunch trains (see the electric field in **b**), the final energy spread generally increases with increased dephasing.

The self-correction mechanism is also robust when other important wakefield effects are taken into account. One such effect is *dephasing*—a backward slippage bringing the driver closer to the accelerating bunch. This is important for laser drivers due to their subluminal speed[26], but also for beam drivers in the case of head erosion[27]. Dephasing leads to an averaging of wakefields over a range of phases, resulting in a change in effective beam loading. Figure 4 shows simulations with varying levels of dephasing per stage (see Methods). For each setting, the self-correction mechanism produces a different current profile: in the presence of dephasing, the equilibrium current profile is a train of microbunches. This occurs because the wakefield cannot be flattened on the scale of the dephasing slippage (10 μm in the example in Fig. 4b), leading to a breakup into multiple regions of local synchrotron oscillation—each individually damped during acceleration, thereby resulting in multiple microbunches (four bunches spaced by 10 μm in Fig. 4b).

Ultimately, the self-correction mechanism will require achromatic and emittance-preserving staging optics. One possible solution includes the use of dipoles and transversely tapered plasma lenses[28]—nonlinear beam optics that enables compact, local chromaticity correction[29]. Having identified a working staging-optics solution, effects that couple the longitudinal and transverse phase spaces can be investigated. This includes ion motion, which causes longitudinally nonuniform focusing and transversely nonuniform acceleration, as well as coherent synchrotron radiation (CSR) in the chicanes. These effects will effectively perturb the wakefield, altering the equilibrium current profile. While such coupling introduces a great deal of complexity, it also introduces new possibilities. For example, at high energy, betatron oscillations cause radiation damping of the transverse emittance, but also produces an energy chirp[29]. This chirp can probably be corrected by the multistage self-correction mechanism at the cost of a slower damping of energy spread. Such an emittance-exchange mechanism could lead to a 6D damping in a linear accelerator similar to that observed in a damping ring—an exciting prospect for applications in high-energy physics. In the shorter term, a proof-of-principle application could be a synchrotron injector, which requires high energy stability and low energy spread, but with relaxed requirements for the transverse emittance.

In conclusion, use of multiple stages enables self-correction mechanisms in plasma accelerators, significantly increasing their tolerance to timing jitter and other beam imperfections. This can have major implications for the realization of plasma accelerators with high beam quality and stability for both high- and low-energy applications.

## References

1. Tajima, T. & Dawson, J. M., Laser electron accelerator. *Phys. Rev. Lett.* **43**, 267–270 (1979).
2. Chen, P., Dawson, J. M., Huff, R. W., & Katsouleas, T., Acceleration of electrons by the interaction of a bunched electron beam with a plasma. *Phys. Rev. Lett.* **54**, 693–696 (1985).
3. Hogan, M. *et al.*, Multi-GeV energy gain in a plasma-wakefield accelerator. *Phys. Rev. Lett.* **95**, 054802 (2005).
4. Leemans, W. *et al.*, GeV electron beams from a centimetre-scale accelerator. *Nat. Phys.* **2**, 696–699 (2006).
5. Joshi, C. & Katsouleas, T., Plasma accelerators at the energy frontier and on tabletops. *Phys. Today* **56**, 47–53 (2003).
6. Leemans, W. & Esarey, E., Laser-driven plasma-wave electron accelerators. *Phys. Today* **62**, 44–49 (2009).
7. Katsouleas, T., Wilks, S., Chen, P., Dawson, J. M. & Su, J. J., Beam loading in plasma accelerators, *Part. Accel.* **22**, 81–99 (1987).
8. Litos, M. *et al.*, High-efficiency acceleration of an electron beam in a plasma wakefield accelerator. *Nature* **515**, 92–95 (2014).
9. Tzoufras, M. *et al.*, Beam loading in the nonlinear regime of plasma-based acceleration. *Phys. Rev. Lett.* **101**, 145002 (2008).
10. Lindstrøm, C. A. *et al.*, Energy-spread preservation and high efficiency in a plasma-wakefield accelerator. *Phys. Rev. Lett.* **126**, 014801 (2021).
11. Steinke, S. *et al.*, Multistage coupling of independent laser-plasma accelerators. *Nature* **530**, 190–193 (2016).
12. Lindstrøm, C. A., Staging of plasma-wakefield accelerators. *Phys. Rev. Accel. Beams* **24**, 014801 (2021).
13. Adolphsen, C. (ed.) *et al.*, International Linear Collider Technical Design Report, Volume 3.I: Accelerator R&D, 2013.
14. Aicheler, M. (ed.) *et al.*, A Multi-TeV linear collider based on CLIC technology: CLIC Conceptual Design Report, 2012.
15. Bonifacio, R., Pellegrini, C. & Narducci, L. M., Collective instabilities and high-gain regime in a free electron laser. *Opt. Commun.* **50**, 373–378 (1984).
16. Schulz S. *et al.*, Femtosecond all-optical synchronization of an x-ray free-electron laser, *Nat. Commun.* **6**, 5938 (2015).
17. Ferran Pousa, A., Martinez de la Ossa, A., Brinkmann, R. & Assmann, R. W., Compact Multistage Plasma-Based Accelerator Design for Correlated Energy Spread Compensation. *Phys. Rev. Lett.* **123**, 054801 (2019).
18. Veksler, V., A new method of acceleration of relativistic particles. *Acad. Sci. U.S.S.R.* **43**, 444 IX (1944).
19. McMillan, E. M., The Synchrotron—A proposed high energy particle accelerator. *Phys. Rev.* **68**, 143–144 (1945).
20. Vincenti, H., Lobet, M., Lehe, R., Vay, J.-L. & Deslippe, J., Exascale Scientific Applications: Scalability and Performance Portability. In *Exascale Scientific Applications: Scalability and Performance Portability* (2018).
21. Panofsky, W. K. H. & Wenzel, W. A., Some considerations concerning the transverse deflection of charged particles in radio-frequency fields. *Rev. Sci. Instrum.* **27**, 967 (1956).
22. Clayton C. *et al.*, Self-mapping the longitudinal field structure of a nonlinear plasma accelerator cavity. *Nat. Commun.* **7**, 12483 (2016).
23. Lu, W., Huang, C., Zhou, M., Mori, W. B. & Katsouleas, T., Nonlinear Theory for Relativistic Plasma Wakefields in the Blowout Regime. *Phys. Rev. Lett.* **96**, 165002 (2006).
24. Dalichaouch, T. N. *et al.*, A multi-sheath model for highly nonlinear plasma wakefields. Preprint at http://arXiv.org/abs/2103.12986 (2021).
25. Lindstrøm, C. A. *et al.*, Staging optics considerations for a plasma wakefield acceleration linear collider. *Nucl. Instrum. Methods Phys. Res. A* **829**, 224–228 (2016).
26. Esarey, E., Schroeder, C. B., & Leemans, W. P., Physics of laser-driven plasma-based electron accelerators, *Rev. Mod. Phys.* **81**, 1229–1285 (2009).
27. Blumenfeld, I. *et al.*, Energy doubling of 42 GeV electrons in a metre-scale plasma wakefield accelerator. *Nature* **445**, 741–743 (2007).
28. Lindstrøm, C. A., Achromatic transport of plasma-accelerated beams using transversely tapered plasma lenses. *To be published* (2021).
29. Raimondi, P. & Seryi, A., Novel final focus design for future linear colliders. *Phys. Rev. Lett.* **86**, 3779 (2001).
30. Michel, P., Schroeder, C. B., Shadwick, B. A., Esarey, E. & Leemans, W. P., Radiative damping and electron beam dynamics in plasma-based accelerators. *Phys. Rev. E* **74**, 026501 (2006).

## Acknowledgements

The author thanks E. Adli, J. Chappell, R. D'Arcy, B. Foster, W. Leemans, J. Osterhoff and K. Põder for helpful comments.

## Methods

**Modelling single-stage nonlinear plasma wakefields.** In the nonlinear regime, plasma wakes can be described by assuming all plasma electrons move in one or more electron sheaths. Building on a model derived by Lu *et al.*[24], which uses a single electron sheath, Dalichaouch *et al.*[25] uses a two-sheath model to more accurately describe the evolution of the electron-sheath radius $r_b$, resulting in the differential equation

$$A(r_b)\frac{d^2 r_b}{d\xi^2} + \frac{B(r_b)}{k_p r_b}\left(\frac{dr_b}{d\xi}\right)^2 + \frac{C(r_b)}{k_p r_b} = \frac{1}{2\pi e c}\frac{I(\xi)}{n_0 k_p r_b^3}, \quad (2)$$

where $k_p$ is the plasma wavenumber, $n_0$ is the plasma density, $\xi$ is the co-moving coordinate, $I(\xi)$ is the current profile, and $c$ and $e$ are the vacuum light speed and electron charge, respectively. The functions $A(r_b)$, $B(r_b)$ and $C(r_b)$, which are all of order unity, are described in full in Ref. 25. These functions depend loosely on several quasi-free parameters, which in this paper are all set equal to those used in Ref. 25: $\Delta_{10} = 1$ and $\Delta_{20} = 3$ represent the normalized initial thicknesses of the two sheaths; $\varepsilon = 0.05$ is the ratio at which an increase in sheath radius increases the inner-sheath thickness (i.e., $\Delta_1 = \Delta_{10} + \varepsilon r_b$); $s = 3$ represents the normalized width of the Gaussian radial density profile of the outer sheath; and $\psi_{min} = -1$ is the approximated minimum vector potential. Solving equation (2), the electric field can be calculated as

$$E_z(\xi) = -D(r_b)\frac{n_0 e}{\epsilon_0} r_b \frac{dr_b}{d\xi}, \quad (3)$$

where $\epsilon_0$ is the vacuum permittivity and $D(r_b)$ is another function of order one (described in Ref. 25). As boundary conditions, we assume that a nondescript laser or beam driver sets up a wake with a maximum electron-sheath radius $R_b$ at $\xi = 0$ (where the electric field is zero, near the middle of the wake), and start the simulation from this point. In this paper, equation (2) is solved numerically to a relative precision of $10^{-7}$.

**Modelling multiple stages separated by magnetic chicanes.** Extending the single-stage acceleration requires the introduction of a new, rudimentary 1D particle-in-cell simulation. Starting with an initial particle distribution in LPS, where each particle has a longitudinal position $\xi$ and an energy $E$, three steps occur at each stage of the multistage accelerator. Firstly, the beam-loaded wakefield is calculated based on the current profile of the particle distribution (a histogram) using equations (2) and (3). Secondly, the particle energy is updated based on their location in the wakefield and the length of the stage $L$,

$$E \mapsto E - e E_z(\xi) L, \quad (4)$$

where $e$ is the electron charge. Thirdly, having exited the stage and traversed a magnetic chicane, the longitudinal position of each particle is updated based on its energy

$$\xi \mapsto \xi + R_{56}\delta, \quad (5)$$

where $R_{56}$ is the longitudinal lattice dispersion of the chicane, and $\delta$ is the relative energy offset of the particle with respect to the nominal energy set point for the chicane $E_0$,

$$\delta = \frac{E}{E_0} - 1. \quad (6)$$

This three-step loop is then repeated for every stage. Use of 1 million macroparticles across 1000 grid cells is found to produce a sufficiently small statistical error.

**Modelling dephasing within accelerator stages.** To model dephasing, each accelerator stage is split into a number of longitudinal sub-steps $N$. In the initial sub-step, the injection phase is shifted backwards by $\Delta\xi/2$, where $\Delta\xi$ is the total longitudinal slippage due to dephasing in a single stage. Subsequently, in each sub-step, the bunch moves forward in the wakefield by a distance $\Delta\xi/N$, where the wakefield (and its beam loading) is re-calculated based on the current profile using equations (2) and (3). The contribution from each sub-step is then integrated to give an averaged wakefield

$$\langle E_z(\xi)\rangle = \frac{1}{N}\sum_{i=0}^{N-1} E_z\left(\xi + \left(\frac{i}{N-1} - \frac{1}{2}\right)\Delta\xi\right). \quad (7)$$

In the calculation of Fig. 4, the size of each sub-step is given by $0.02/k_p$, where the number of sub-steps in each stage is rounded up to the nearest integer (and imposing a minimum of 10 sub-steps).